# Knowledge spillovers: does the geographic proximity effect decay over time? A discipline-level analysis, accounting for cognitive proximity, with and without self-citations[1]

Giovanni Abramo
*Laboratory for Studies in Research Evaluation*
*at the Institute for System Analysis and Computer Science (IASI-CNR)*
*National Research Council of Italy*
ADDRESS: Istituto di Analisi dei Sistemi e Informatica, Consiglio Nazionale delle Ricerche, Via dei Taurini 19, 00185 Roma - ITALY
giovanni.abramo@uniroma2.it

Ciriaco Andrea D'Angelo
*University of Rome "Tor Vergata" - Italy and*
*Laboratory for Studies in Research Evaluation (IASI-CNR)*
ADDRESS: Dipartimento di Ingegneria dell'Impresa, Università degli Studi di Roma "Tor Vergata", Via del Politecnico 1, 00133 Roma - ITALY
dangelo@dii.uniroma2.it

Flavia Di Costa
*Research Value s.r.l.*
ADDRESS: Research Value, Via Michelangelo Tilli 39, 00156 Roma- ITALY
flavia.dicosta@gmail.com

**Abstract**
This work analyzes the variation over time of the effect of geographic distance on knowledge flows. The flows are measured through the citations exchanged between scientific publications, including and excluding self-citations. To calculate geographic distances between citing and cited publication, each publication is associated with a "prevailing" territory, according to the authors' affiliations. We then apply a gravity model to account for the research size of the territories, in terms of cognitive proximity of citing-cited publications. The field of observation is the 2010-2017 world publications citing the 2010-2012 Italian publications, as indexed in the Web of Science. The results show that in domestic knowledge flows, geographic proximity remains an influential factor through time, although with differences among disciplines and trends of attenuating effects. Finally, we replicate the analyses of knowledge flows but with the exclusion of self-citations: in this manner the effect of geographic proximity seems reduced, particularly at the national scale, but the differences (with vs without self-citations) lessen through time. As shown in previous works, the effect of distance on continental flows is modest (imperceptible for intercontinental flows), yet here too time has some influence, including concerning exclusion of self-citations.

# 1. Introduction

The speed and diffusion of knowledge flows determine the pace of scientific progress. The understanding of the geographic and temporal dimensions of the general diffusion process, but also that within the different fields, would be of great assistance in modeling the progress of science, as well as in practical support for research policies.

Given that the factor of geographic distance between new knowledge producers and potential users has long been considered to have an effect on knowledge flows, this has been one area of particularly extensive investigation. Starting from the assumption that citation linkages between publications imply a flow of knowledge from cited to citing authors (Mehta, Rysman & Simcoe, 2010; Van Leeuwen & Tijssen, 2000), a series of empirical studies have demonstrated that geographic proximity favors knowledge flows among scholars (Matthiessen, Schwarz & Find, 2002; Börner, Penurnarthy, Meiss & Ke, 2006; Liu & Rousseau, 2010; Pan, Kaski & Fortunato, 2012).

In recent years, some scholars have developed a much broader and more complex concept of space than the purely geographic one, as a milieu for knowledge spillovers. This space contemplates different types of proximity (geographic, cognitive, social, institutional) that become manifest and interact within it (Boschma, 2005; Capello, 2009; Breschi & Lissoni, 2001; Wuyts, Colomb, Dutta & Nooteboom, 2005; Tranos & Nijkamp, 2013; Capello & Caragliu, 2018; Ahlgren, Persson & Tijssen, 2013; Frenken, Hardeman & Hoekman, 2009).

Wuestman, Hoekman and Frenken (2019) in their study of the geography of citations, analyzed the role of organizational, regional and national co-location on the probability of citation for 5.5 million pairs of articles. They demonstrated that, in particular, the geographic bias in citations is actually weak when cognitive proximity (i.e. sharing similar knowledge bases) between citing-cited publications is taken into account.

Balland, Boschma and Frenken (2015) proposed an extension of the proximity framework that accounts for the dynamics of co-evolution between knowledge networking and proximity. For each proximity dimension, they analyze how proximities might increase over time as a result of past knowledge ties. These dynamics are captured through the processes of learning (cognitive proximity), integration (organizational proximity), decoupling (social proximity), institutionalization (institutional proximity), and agglomeration (geographic proximity).

The aspect of the sensitivity of the geographic proximity effect to the time factor has as yet been little studied. Wang and Zhang (2018) examined the diffusion of knowledge over time, recurring to citing-cited publications. Using a set of research articles in physics, the study compared the number of citations from domestic scholars to those from scholars abroad, in each year after publication. Results show that domestic citations accumulate faster and reach their peak earlier than foreign ones, confirming that the diffusion of knowledge is influenced by spatial boundaries, although to a decreasing extent as time passes from the date of publication. The authors also stressed some limits of the study: i) the presence of the self-citation phenomenon, which affects domestic citing of publications (Aksnes, 2003); ii) the difficulty of generalizing the results, since they deal with the field of physics, characterized by a strong rate of internationalization, meaning that in other fields there could well be stronger effects from national boundaries (Wagner, Whetsell & Leydesdorff, 2017); iii) the fact that the analysis is based on publications indexed in the Web of Science (WoS), which is biased against non-English publications, meaning that the effect of national boundaries on knowledge spillover could be underestimated (Van Leeuwen, Moed, Tijssen, Visser & Van Raan, 2001).

Fontana, Montobbio and Racca (2019) focused on the knowledge flows in economics, investigating how the probability of citation is affected by time since publication, and the geographic location and scientific topic of each paper. Results revealed four overlapping effects: i) there is a "home bias" effect in citations (for example a publication originating in Europe is 39% more likely to get a citation from a random European publication than from a random USA publication); ii) this effect fades over time; iii) USA publications retain a longer lasting world impact vis-à-vis other countries; iv) there is a higher speed of diffusion and faster obsolescence in the United States, i.e. citations in the USA arrive more rapidly and show a higher rate of decay.

The present study aims to better understand the interaction between geographic proximity and time in determining the spread of knowledge dissemination. In particular, it investigates the extent to which the effect of geographic proximity in favoring knowledge flows, tends to fade over time. Starting from the notion that citations reflect acknowledgement of cognitive content, in determining the distance of knowledge flows, i.e. between the territories of the citing and cited publication pairs, we try to limit the bias of cognitive proximity. In fact, because cognitively related knowledge may be geographically concentrated, the probability that two territories cite each other's publications depends not only on the size of their research (number of publications), but also on the cognitive proximity of their research.[2] The study goes beyond the sectoral level of previous works, examining the relationship between knowledge flows and geographic distance as a function of time at both the overall level of the sciences and at the level of the single disciplines, both including self-citations and excluding them. In fact, there is no full agreement among scholars on whether to include or not self-citations in the analyses of the effect of distance on knowledge flows. A self-citation represents a peculiar flow of knowledge "within" the same author, rather than among authors, and inevitably amplifies the effect of geographic proximity.

There are a number of reasons leading us to expect differences in the influence of geographic proximity across the fields of research. First, the citation behavior of authors differs across fields (Hurt, 1987; Vieira & Gomes, 2010). Second, the research topics of certain fields could be more territory specific, addressing local needs, and so with more localized spillovers. Third, when included in the analysis, self-citations amplify the effect of geographic proximity (Aksnes, 2003), and it is known that self-citation rates vary across fields (Ioannidis, Baas, Klavans & Boyack, 2019). Finally, field-focused research organizations may be more or less clustered geographically, and the clusters could also differ in their size and internal numerosity. Therefore, to the extent that scientists of places concentrate their research efforts on certain topics (Boschma, Heimeriks & Balland, 2014), citations reflecting intellectual recognition will also be more geographically concentrated (Head, Li and Minondo, 2019). Hence, in principle, the geographic proximity effect in citations could be fully explained by the geographic concentration of specialized knowledge.

The analyses will evaluate the effect of geographic proximity, accounting for the cognitive proximity between citing and cited publications. The method to take into account cognitive proximity, presented in the next section, has been conceived by the same authors (Abramo, D'Angelo, & Di Costa, 2020b), and represents an innovation in the literature on the subject.

---

[2] A paper in biology is more likely to be cited by one in biology or clinical medicine than by one in architecture

This empirical investigation continues from two previous works by the authors (Abramo, D'Angelo, & Di Costa, 2020a; Abramo, D'Angelo, & Di Costa, 2020b), analyzing the influence of geographic proximity on knowledge flows, as measured through citations between scientific publications. In all of these works, the reference is to the specific case of knowledge flows generated in Italy.

The previous analyses, conducted at the level of individual disciplines, demonstrate that despite the diffusion of information and communication technology, geographic distance remains an influential factor in the process of knowledge flows between territories. The analyses, replicated at three different geographic scales (national, continental and intercontinental), have shown that the effect from "geographic distance" is mostly confined to the domestic scale, with some lesser influence at the continental scale (Europe). For greater distances (intercontinental scale) the influence fades to the point of irrelevance, meaning that the citation process experiences no effect from the factor of distance.

This work, while accounting for cognitive proximity, expands the previous analyses by introducing the time dimension. In detail, the analysis aims to answer the following research questions. Given the geographic distance effect on knowledge flows between territories:

i) What is its variation over time?

ii) What are the differences in the effect between disciplines, at national, continental, and intercontinental scales?

iii) What biases do self-citations induce, and do these too vary with time?

To conduct our investigation, we analyze the 2010-2017 world publications citing the 2010-2012 Italian publications (citation time window ranges between 5 and 8 years), as indexed in the Clarivate Analytics Italian national citation report (I-NCR), extracted from Web of Science (WoS) core collection.

In the next section, we present the methods and the data of analysis. Section 3 reports and discusses the results; Section 4 offers concluding remarks.

## 2. Methods and data[3]

To test the influence of geographic proximity on knowledge flows, we apply a gravity model similar to that used by Ponds, Van Oort and Frenken (2007) for the study of scientific collaborations between different types of institutions. The gravity model is derived from the "law of universal gravitation" proposed by Newton in 1687, stating that the gravitational force between masses decreases with the distance between them, according to an inverse-square law. In the economics, gravity models are commonly used to explain international trade: bilateral trade between two countries is proportional to their economic mass (i.e. GDP or population) and inversely related to their geographical distance.

The gravity model adopted here is based on two assumptions:

---
[3] The description of methods and data aligns with the one previously presented by the authors (Abramo, D'Angelo, & Di Costa 2020a), and some text passages have been retained.

- the flow of knowledge between any two territories can be measured through the citing publications authored by research organizations in the first territory, and the relevant cited publications authored by the research organizations in the second;
- citations between two territories increase with the research size (in terms of publications bearing cognitive proximity) of both, and decrease with the distance between them.

Various approaches could be envisaged to assign the publications to geographic entities:

i) to each of the territories of the institutions in the address list;
ii) to one single territory, by the frequency of authors (or institutions) of the territory in the address list, or by the affiliation of the corresponding author, or by the affiliation of the first and last authors in non-alphabetically ordered bylines;
iii) by fractionalizing the publication by the number of territories, institutions or authors.

We determined to adopt two distinct conventions for the cited and citing publications, as in Abramo, D'Angelo and Di Costa (2020a), to which we refer the reader for a thorough discussion:

- For cited publications, we define a publication as "made in" a territory if the majority of its co-authors are affiliated to organizations located in that territory.[4]
- Differently from the cited publications, for the citing publications the I-NCR reports only the address list without the link to authors. We define then a publication as "made in" a territory if the majority of its addresses refers to that territory.[5]

From an operational point of view, the preparatory work for the elaboration required three steps: i) construction of the dataset, consisting of the pairs of cited and citing publications; ii) assignment of the geographic attribute to each cited publication and the relevant citing ones; iii) calculation of the geographic distances between citing and cited publications.

The Clarivate Analytics Italian national citation report (I-NCR) registers all publications with "Italy" in the affiliation list. Let P denote the set of the cited publications indexed in such report. For each publication in P, we reduce all addresses to city + country expressions (e.g. "Rome, Italy"). Each "city" is then matched to the corresponding LAU level (local administrative unit, 11107 in all)[6], using the official lists of the National Institute of Statistics (ISTAT).[7]

Overall, 255,399 publications are registered in the 2010-2012 I-NCR, of which 184,177 had received at least one citation by the close of 2017. 161,680 were assigned to an Italian LAU (the remaining 22,497 publications with no prevalent territory are excluded from the analysis), and had received 3,002,835 total citations from 1,800,037 unique citing publications.

---

[4] In case of multiple affiliations, we adopt a fractional counting method. Territories listed in bibliometric addresses of an author with "m" different affiliations, account for 1/m each.

[5] This convention has some obvious limits: a citing publication could be attributed to a given territory when in fact the authors from that territory did not reach a "majority" within the byline; the full counting of each of the authors' addresses distorts the result in the presence of authors with multiple affiliations; finally, the corresponding author ends up having twice as much weight as the others, for the simple fact that their affiliation appears twice in the address list. In order to evaluate the effect of such limits, we extracted a random sample of 1,000 cited publications from the dataset and, for each citing record of such publications (17,216 in all), we downloaded the author-affiliation field by means of the "Advanced Search" interface in the online WoS portal. The application of both conventions to such set of citing publications reveals that in 96.8% of cases the "made in" territory remains the same.

[6] The LAU level consists of municipalities or equivalent units in the 27 EU member states.

[7] https://www.istat.it/it/archivio/6789, last accessed on 22 May, 2020.

The analysis of knowledge flows will be carried out at three distinct geographic scales:
- the national one, in which the citing publications assigned to "Italy" are attributed to one and only one LAU of the Italian territory, always on the basis of the prevalence criterion;
- the international one, where the citing publications will be attributed to one and only one country on the basis of the prevalent NUTS0 code;[8] we will distinguish also between the continental (Europe) and the intercontinental (extra-Europe) context.

We then measure the "distances" of the citation flows, along the geodetic line[9] that joins the prevalent Italian LAU of production of the aforementioned publication with:
- the citing Italian LAU, for national analysis,
- the capital of the citing country, for international analysis.

Geographical coordinates of points of interest were extracted from ISTAT for Italian LAUs,[10] and from Simplemaps® for capitals of citing countries.[11] Starting from the longitude and latitude coordinates of each point, we calculate the geodetic distance between territories as the shortest distance on a sphere.[12]

In this work, we account for the cognitive proximity of the citing-cited publications. Previous studies measured the research size of the citing territory by the total number of publications made in that territory (Pan, Kaski, & Fortunato, 2012), or solely by the number of publications falling in the same field as the cited publication (Abramo, D'Angelo, & Di Costa, 2020a). Here, we adopt a different method to measure the research size of the citing territory, applied for the first time by Abramo, D'Angelo and Di Costa (2020b). The underlying rationale is that the real research size of the citing territory consists of the publications cognitively close to the cited ones, which are less than the total publications but more than solely those falling in the same field as the cited ones.

We measure the research size of the territory of the cited publication by the total number of publications of that territory falling in the same WoS subject category (SC).

Much more complex is the way we measure the research size of the territory of the citing publications. Citing publications may fall or not fall in the same SC as the cited publication. We first calculate the SC frequency distribution of all world publications citing all Italian publications within a certain SC. The research size of the territory of each citing publication is the weighted sum of that territory's publications falling in the above identified SCs, whereby the weights correspond to their frequency distribution.

To exemplify, let us assume that we want to measure the knowledge flows generated by the cited publications in Paleontology made in LAU Milan, to LAU Turin. We consider all publications in the dataset falling in the SC Paleontology. Relevant world citing publications in the observed period fall in 93 different SCs:[13] 45% in Paleontology; 18.9% in Geosciences, multidisciplinary; 9.4% in Geology; 8.6% in Geography, physical;

---

[8] The NUTS classification (Nomenclature of territorial units for statistics) is a system subdividing the economic territory of the European Union into hierarchical levels.

[9] In the literature, this method of measuring geographic distance has been adopted in Maurseth and Verspagen, 2002; Broekel and Mueller, 2018; Ahlgren, Persson and Tijssen, 2013; Jiang, Zhu, Yang, Xu & Jun, 2018. Some scholars have instead adopted the travel time between two points (Crescenzi, Nathan, and Rodríguez-Pose, 2016; Ponds, Van Oort & Frenken, 2007).

[10] https://www.istat.it/it/archivio/6789, last accessed on 22 May 2020.

[11] https://simplemaps.com/resources/free-country-cities, last accessed on 22 May 2020.

[12] https://en.wikipedia.org/wiki/Great-circle_distance, last accessed on 22 May 2020.

[13] Papers published in multi-category journals are full counted in each category.

and the remaining 18% are dispersed across the remaining 89 SCs. The research size of Milan is measured by the 2010-2012 cited publications in Paleontology made in Milan (52 in all). The research size of Turin, instead, is measured by the weighted average (where weights are represented by the above percentages) of the 2010-2017 publications made in Turin and falling in the above 93 SCs (189 in all).

The gravity model adopted for the national analysis in each SC is:

$$C_{ij} = k \cdot \frac{M_i^\alpha M_j^\beta}{d_{ij}^\gamma}$$

[1]

with:

$C_{ij}$ = number of citations to publications made in LAU $i$ by the publications made in LAU $j$
$k$ = gravity constant
$M_i$ = total number of publications made in LAU $i$ in the 2010-2012 period
$M_j$ = weighted number of publications made in LAU $j$ in the 2010-2017 period
$d_{ij}$ = geodetic distance between cited LAU $i$ and citing LAU $j$

For the international analysis, the following distinctions apply:

$C_{ij}$ = number of citations to publications made in LAU $i$ by the publications made in country $j$
$M_j$ = weighted number of publications made in country $j$ in the 2010-2017 period
$d_{ij}$ is the distance between cited LAU $i$ and the capital of the citing country $j$

Applying a logarithmic transformation to all variables of equation [1], we obtain:

$$\ln(C_{ij}) = \ln(k) + \alpha \ln(M_i) + \beta \ln(M_j) - \gamma \ln(d_{ij}) + \varepsilon$$

[2]

The coefficients of a log-log model represent the elasticity of the Y dependent variable with respect to the X independent variable. For example, for the distance variable ($d_{ij}$) an elasticity of one ($\gamma = 1$) indicates that a 1% increase in the distance is associated with a 1% decrease in citations exchanged, on average.

For the 2010-2012 triennium, the I-NCR dataset contains 255,399 Italian publications, 184,177 of which had received at least one citation up to the close of 2017. 161,680 were assigned univocally to an Italian LAU,[14] and had received 3,002,835 total citations from 1,800,037 citing publications. The overall dataset was broken down by SC (244 in all, according to the WoS classification schema) of the hosting journal.[15] In turn, the SCs are grouped in disciplinary areas (DAs, six in all), as defined by the OECD, applying a category-to-category mapping available on the Incites-Clarivate Analytics portal.[16]

---

[14] The remaining publications had no prevalent LAU, and have been assigned to none.
[15] Publications in multi-category journals are assigned to each category.
[16] http://help.prod-incites.com/inCites2Live/5305-TRS.html, last accessed on 22 May 2020.

## 3. Analysis

The following three subsections provide: i) an analysis of the average distances of real citation flows as a function of time, by disciplinary and geographic scale; ii) a representation of the time pattern of geographic proximity, accounting for the research size of territories and cognitive proximity of research; and iii) a study of the effects of self-citations on the previous analyses.

In all cases, we consider three different geographic entities: national, European and non-European, depending on the location of the citing publications.

### 3.1 Descriptive analysis at disciplinary area level

Figure 1 shows, for each geographic scale, with no distinctions for disciplinary areas (i.e. overall dataset), the average distances between the territories of origin of the cited and cited publication, in function of the time elapsed between the date of citing publication and the date of cited publication. Time 0 indicates that the citing publication is published in the same year as the cited one. The trends in national and EUR flows show increasing distances over time as opposed to extra-EUR flows, which show a curvilinear trend with a maximum of three years. For national flows, the average distance is 109 km, between citing and cited publications of the same year. When the citing work is published 7 years after cited one, the average distance almost doubles (191 km).[17]

*Figure 1: Time pattern for average distances of citation flows: overall dataset*

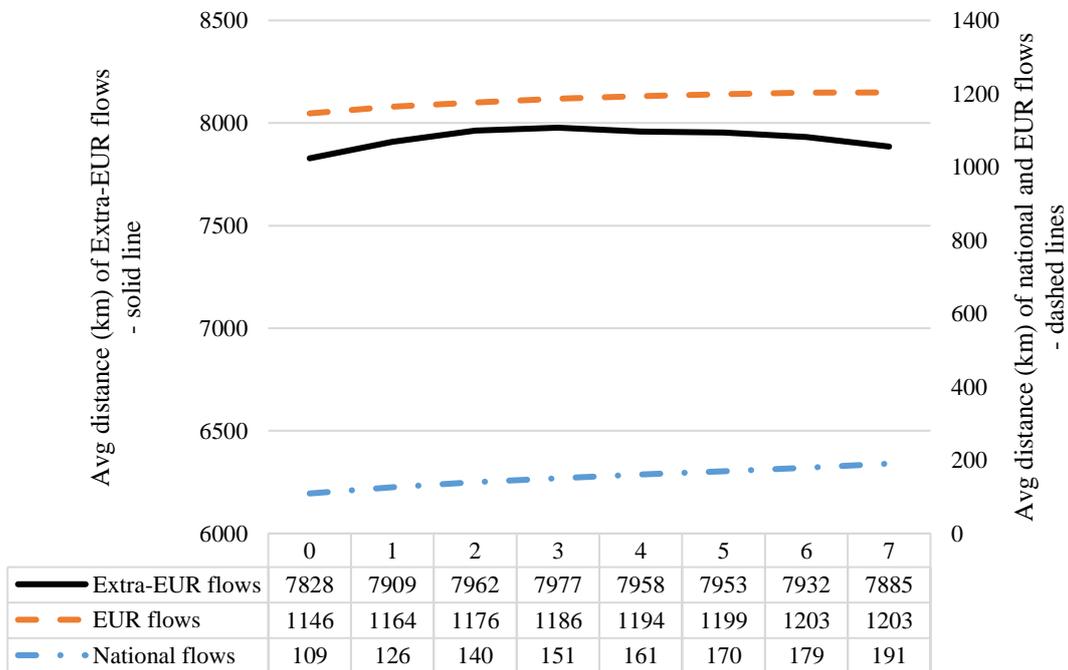

Citation delay: time elapsed between citing and cited publications dates (years)

---

[17] In Italy, the maximum possible geographic distance between two cited-citing LAUs is 1119 km from south to north, i.e. Catania to Aosta.

Figure 2 shows the trends of the average distances in the national context, in each DA. In all DAs, average distances are always below 300 km. The trends over time are increasing in all DAs, indicating a progressive reduction of the proximity effect as years pass. In detail, the highest distances are found in three DAs: Humanities (t = 0, 3, 4, 7); Medical and health sciences (t = 1, 2) and Agricultural sciences (t = 5, 6). The lowest values are in Engineering and technology, at all times except t = 7 (Natural sciences).

Figure 3 shows the trends of average distances of knowledge flows within Europe. The values remain below the threshold of 1350 km.[18] We observe that in Medical and health sciences the distance vs time pattern is always increasing, and almost always in Natural sciences and Engineering and technology. In other DAs, the values oscillate with time.

Figure 4 shows the results of the same analysis conducted at the extra-EUR level. The average distance values are always below a threshold of 8400 km.[19] For almost all disciplines, the trends show decreasing values over the final two years. Engineering and technology presents the singular case of a constantly decreasing trend. Humanities instead demonstrate a more irregular and oscillating trend, with major percentage variations between periods.

*Figure 2: Time pattern of the average distance of national knowledge flows, by discipline*

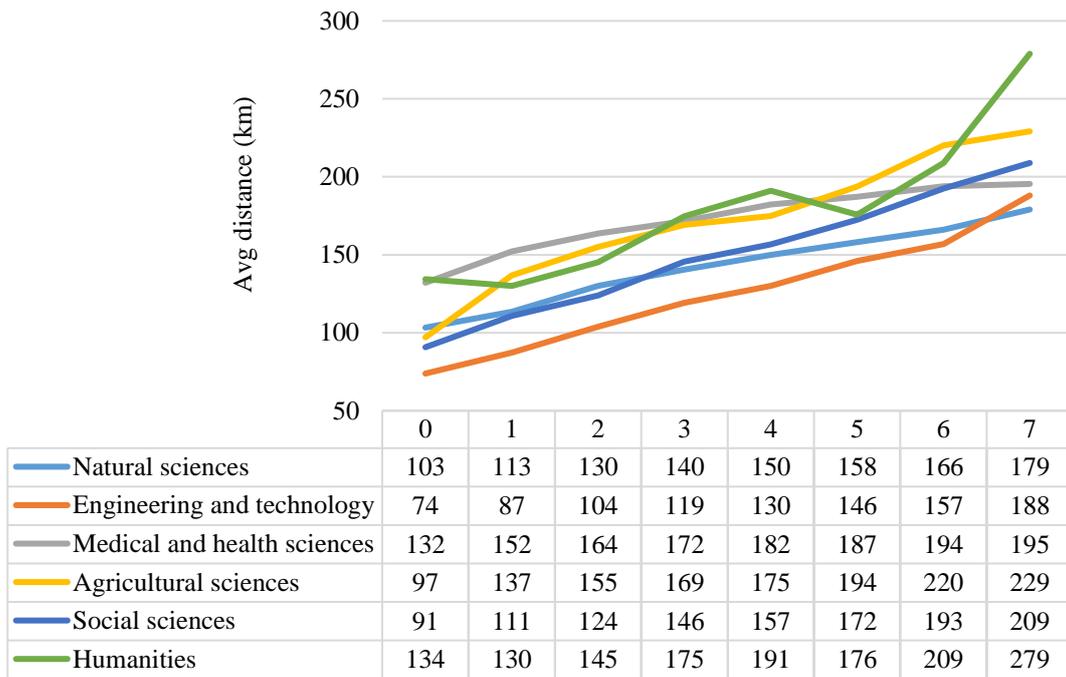

| | 0 | 1 | 2 | 3 | 4 | 5 | 6 | 7 |
|---|---|---|---|---|---|---|---|---|
| Natural sciences | 103 | 113 | 130 | 140 | 150 | 158 | 166 | 179 |
| Engineering and technology | 74 | 87 | 104 | 119 | 130 | 146 | 157 | 188 |
| Medical and health sciences | 132 | 152 | 164 | 172 | 182 | 187 | 194 | 195 |
| Agricultural sciences | 97 | 137 | 155 | 169 | 175 | 194 | 220 | 229 |
| Social sciences | 91 | 111 | 124 | 146 | 157 | 172 | 193 | 209 |
| Humanities | 134 | 130 | 145 | 175 | 191 | 176 | 209 | 279 |

Citation delay: time elapsed between citing and cited publications dates (years)

---

[18] As a reference, the distance between Rome and London is 1,434 km.
[19] The Rome-New York distance, for example, is 6,891 km; Rome-Tokyo is 9,874 km; Rome-Beijing 8,139 km.

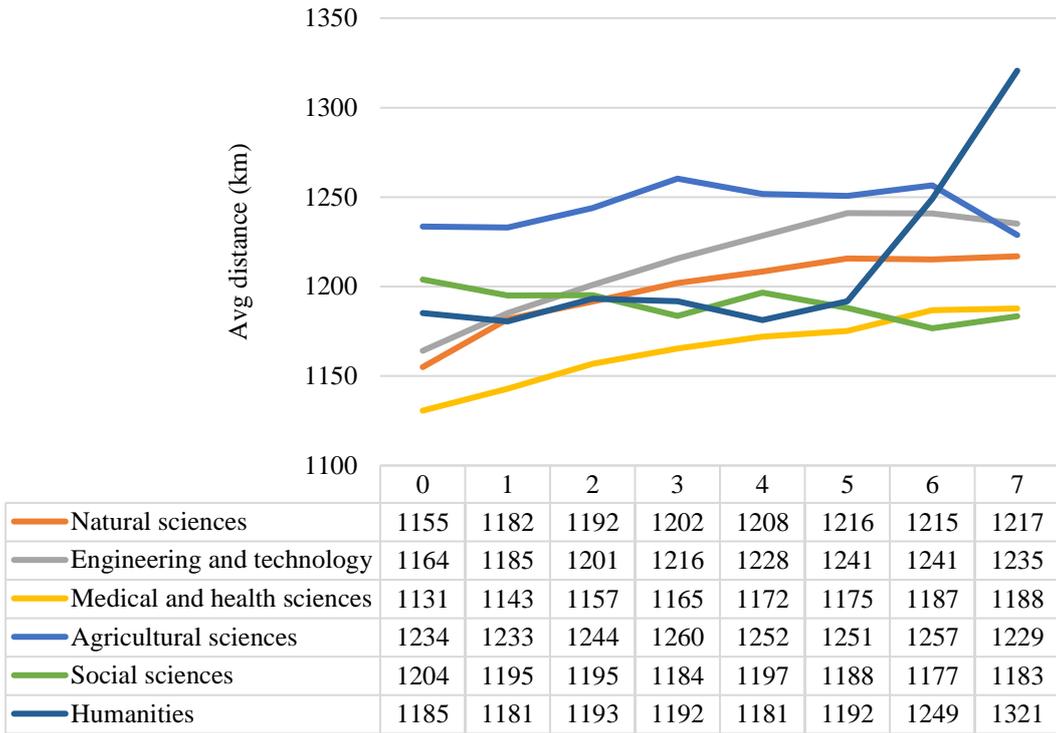

*Figure 3: Time pattern of average distance of knowledge flows in Europe, by discipline*

| | 0 | 1 | 2 | 3 | 4 | 5 | 6 | 7 |
|---|---|---|---|---|---|---|---|---|
| Natural sciences | 1155 | 1182 | 1192 | 1202 | 1208 | 1216 | 1215 | 1217 |
| Engineering and technology | 1164 | 1185 | 1201 | 1216 | 1228 | 1241 | 1241 | 1235 |
| Medical and health sciences | 1131 | 1143 | 1157 | 1165 | 1172 | 1175 | 1187 | 1188 |
| Agricultural sciences | 1234 | 1233 | 1244 | 1260 | 1252 | 1251 | 1257 | 1229 |
| Social sciences | 1204 | 1195 | 1195 | 1184 | 1197 | 1188 | 1177 | 1183 |
| Humanities | 1185 | 1181 | 1193 | 1192 | 1181 | 1192 | 1249 | 1321 |

Citation delay: time elapsed between citing and cited publications dates (years)

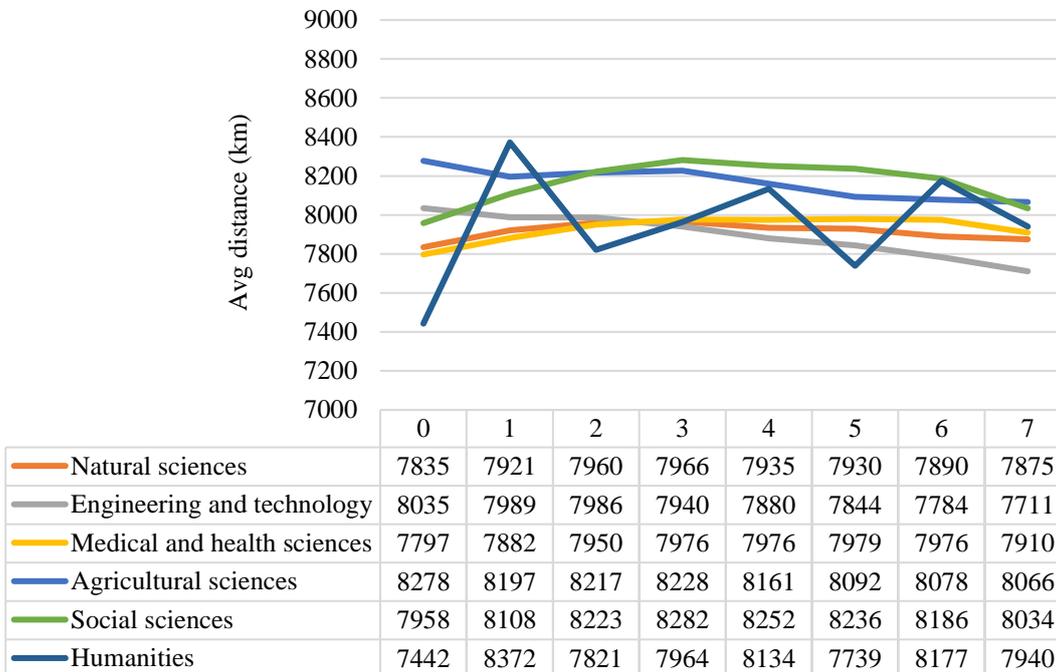

*Figure 4: Time pattern of average distance of intercontinental knowledge flows, by discipline*

| | 0 | 1 | 2 | 3 | 4 | 5 | 6 | 7 |
|---|---|---|---|---|---|---|---|---|
| Natural sciences | 7835 | 7921 | 7960 | 7966 | 7935 | 7930 | 7890 | 7875 |
| Engineering and technology | 8035 | 7989 | 7986 | 7940 | 7880 | 7844 | 7784 | 7711 |
| Medical and health sciences | 7797 | 7882 | 7950 | 7976 | 7976 | 7979 | 7976 | 7910 |
| Agricultural sciences | 8278 | 8197 | 8217 | 8228 | 8161 | 8092 | 8078 | 8066 |
| Social sciences | 7958 | 8108 | 8223 | 8282 | 8252 | 8236 | 8186 | 8034 |
| Humanities | 7442 | 8372 | 7821 | 7964 | 8134 | 7739 | 8177 | 7940 |

Citation delay: time elapsed between citing and cited publications dates (years)

## 3.2 Gravity model

The gravity model accounts for the research size of the territories, and the cognitive proximity of the cited-citing pairs. The temporal dimension is inserted by measuring citation flows as a function of the citation time window.[20] Applying the model, figures 5 and 6 present the estimates of coefficients for the national and continental (EUR) flows, calculated by OLS.[21]

Figure 5 shows the results at the national scale: for all disciplines, the coefficients $\gamma$ estimated for the regressor $d_{ij}$ (lg $d_{ij}$) present a decreasing trend as the citation time window increases. Although with different intensities, the values are all negative, indicating that, accounting for the research size and cognitive proximity, as distance between domestic territories increases, there is on average a decrease in knowledge flows. This confirms the presence of geographic proximity effect, but also indicates that the effect stabilizes over time. Among the disciplines, Natural science shows the highest values in absolute terms, indicating a greater influence from the geographic factor compared to the one for Humanities.

*Figure 5: Gravity model applied to national flows: coefficient of the natural log of distance with variation of time (i.e. citation window)*

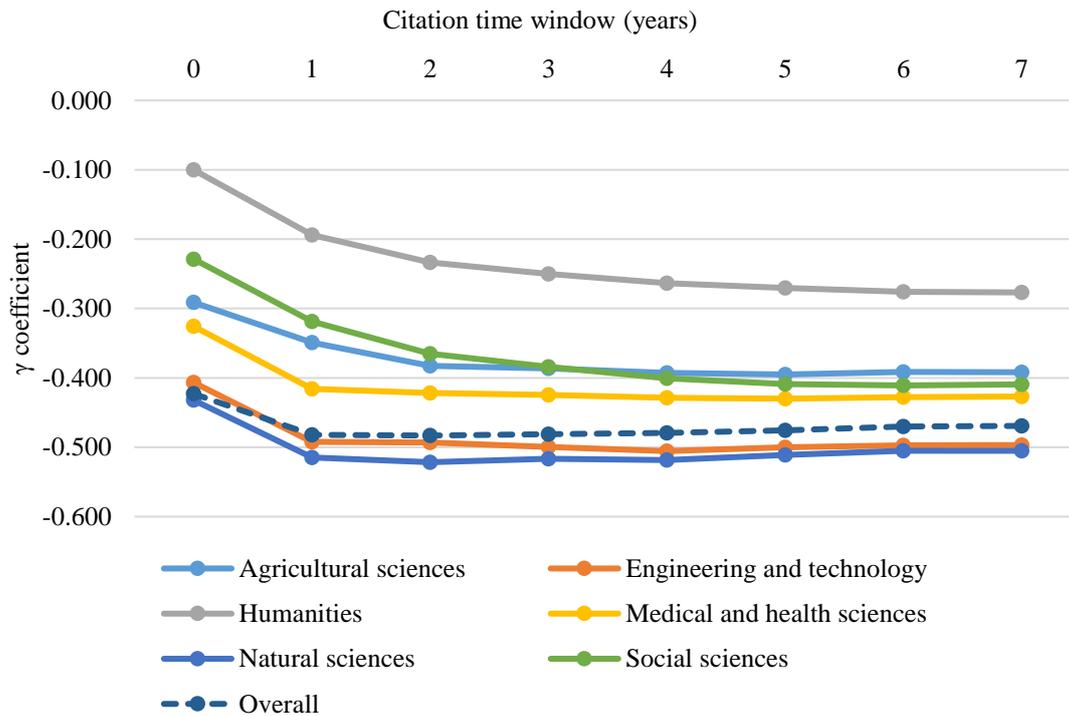

---

[20] In the analyses presented in the previous Section, the time dimension was represented by the difference between the years of the cited and citing publications. In the proposed analysis, instead, the citation time window will be considered and the dependent variable of the model will measure the cumulative citation flows within the time window and not referred to a single year. This allows for a sufficiently large number of observations to robustly estimate OLS models.

[21] The analysis of Extra-EUR flows is not presented because the $\gamma$ coefficients are never statistically significant.

Figure 6 shows the results from the same analysis at the continental scale. Overall, as well for the individual areas of Engineering and technology, Natural sciences and Medical and health sciences, the γ coefficients show increasing trends over the entire period of increasing citation window, indicating a progressive decrease in the effect of geographic proximity on knowledge flows. Interestingly, in 10 of the 56 estimated OLS models (combination of 7 fields and 8 citation windows), the distance-time coefficient is not significant. Moreover, compared to the national case, the trends with increase of time are more erratic.

*Figure 6: Gravity model applied to continental flows: coefficient of distance with variation of time (citation time window)*

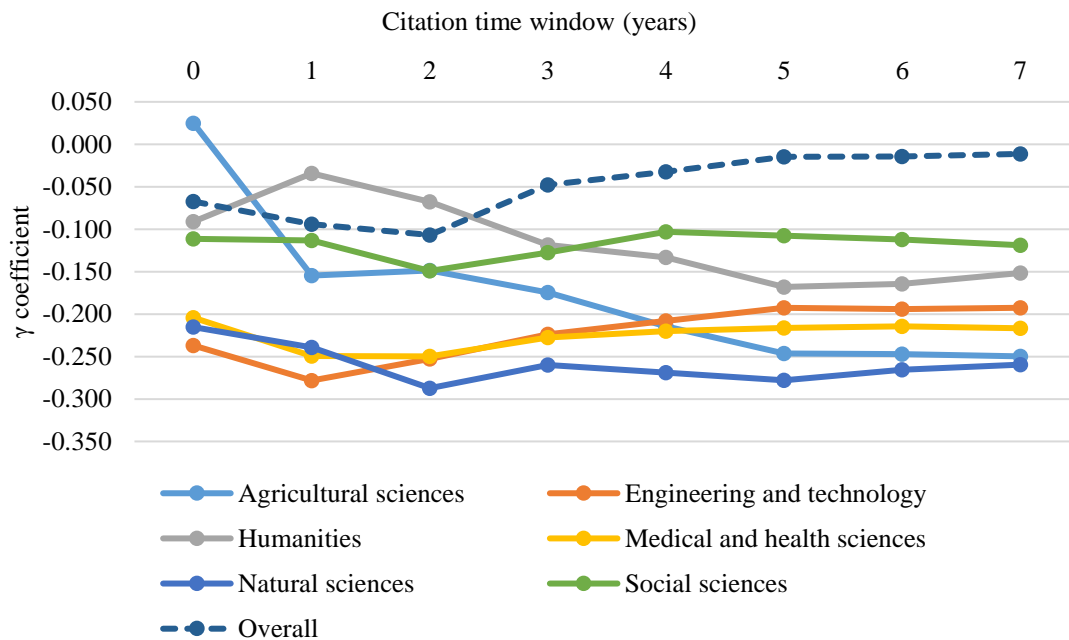

### 3.3 The effect of self-citations

In this subsection we measure the potential impact of the bias introduced by self-citations (Wang & Zhang, 2018; Abramo, D'Angelo, & Di Costa, 2019). Aksnes (2003) had emphasized that in the case of Norway "self-citations contribute to a relatively large share of the citations. In the standard indicators based on a three-year citation window on average more than one third of the citations represent author self-citations. However, this share is significantly decreasing when citations are traced for longer periods."

For the analysis, we first obtain a dataset derived from the original one, "purified" of the contribution from self-citations. Using this, Figure 7 then replicates Figure 1, showing the average distance of the citation flows as time varies, at the overall scientific level. The trends are similar to those of the complete dataset of Figure 1, although the values of distances using the purified data are always greater, for all geographic scales and times. The percentage variations of average distance in the "including self-citations" vs "excluding self-citations" comparison are substantial only for national flows, in the range of 72% (t = 7) to 177% (t = 0). At t = 0, the contribution of self-citations serves to triple the value of average distances (109 vs 302 km). As expected, the weight of self-citations drops off drastically at broader geographic scales: the percentage variations never exceed 1.2%, no matter the time delay.

*Figure 7: Time pattern of the average distance of citation flows, at an overall level: dataset without self-citations*

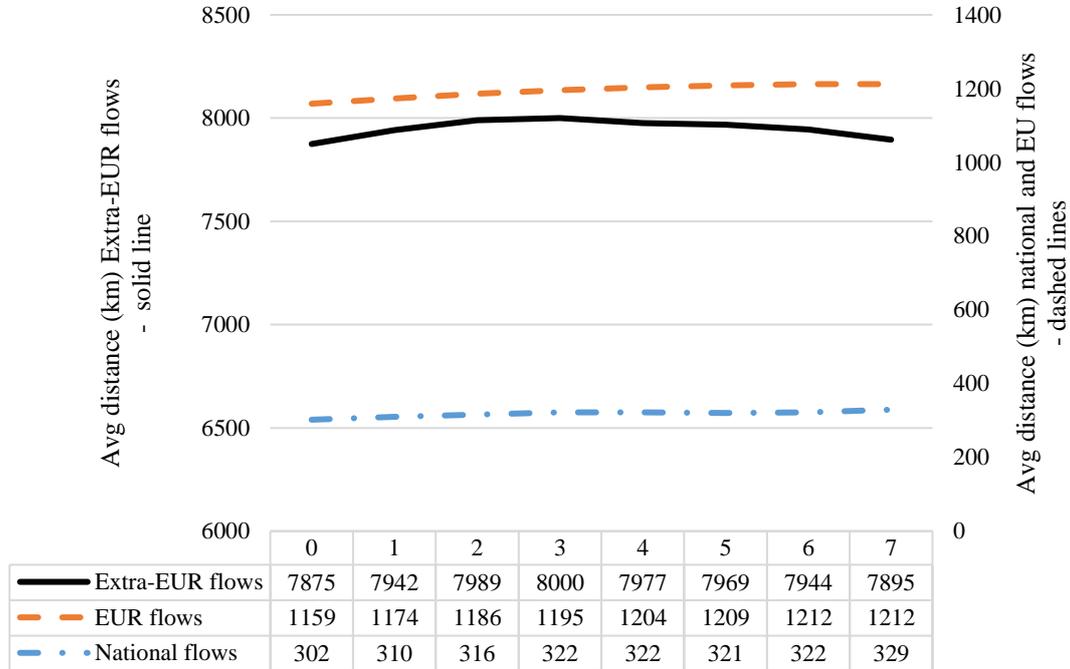

Citation delay: time elapsed between citing and cited publications dates (years)

Figure 8 shows the values of the average distance of national citation flows over time, excluding self-citations, at the level of the individual DA. Here too, without self-citations, the values of average distance are always greater, regardless of geographic dimension or time. The trends are also increasing with time, although this not as evident as with the inclusion of self-citations (Figure 3).

Comparing "with" vs "without self-citations", the greatest percentage variations of average distance are in Engineering and technology (any $t$ but $t = 7$), with values decreasing in the range 316% to 73%. The smallest percentage changes are mostly in Humanities ($t = 0, 3, 4, 5, 7$), then Medical and health sciences (delays 1, 2) and Social sciences (delay 6). The increase in average distance is more pronounced, for all $t$ considered, in the DAs characterized by greater citation intensities, and so a probably greater share of self-citations.

Figure 9 shows the estimates for the gravity model applied to national flows, excluding self-citations, with variation in the citation time window. The trends of the estimates of the $\gamma$ coefficients of the variable $d_{ij}$ (lg $d_{ij}$), for each DA, show trends similar to those of the analysis on the complete dataset (Figure 5). Here, the curves representing the time pattern of the $\gamma$ coefficient never intersect, indicating a stratification of the disciplines: all values are within the range -0.6 to -0.1. The higher values (i.e. stronger proximity effect) are in the scientific-technological areas; lower ones are in the Social Sciences and Humanities, with Medical science and Agricultural sciences representing an intermediate band. The trend of the $\gamma$ coefficient, in which the curves progressively flatten, indicates that the geographic proximity effect stabilizes (i.e. progressive saturation) over time, in all DAs. In other words, in the year of the publication and those immediately following, the proximity effect is important, but it attenuates over time.

*Figure 8: Time pattern of the average distance of citation flows at national level, by discipline: dataset without self-citations*

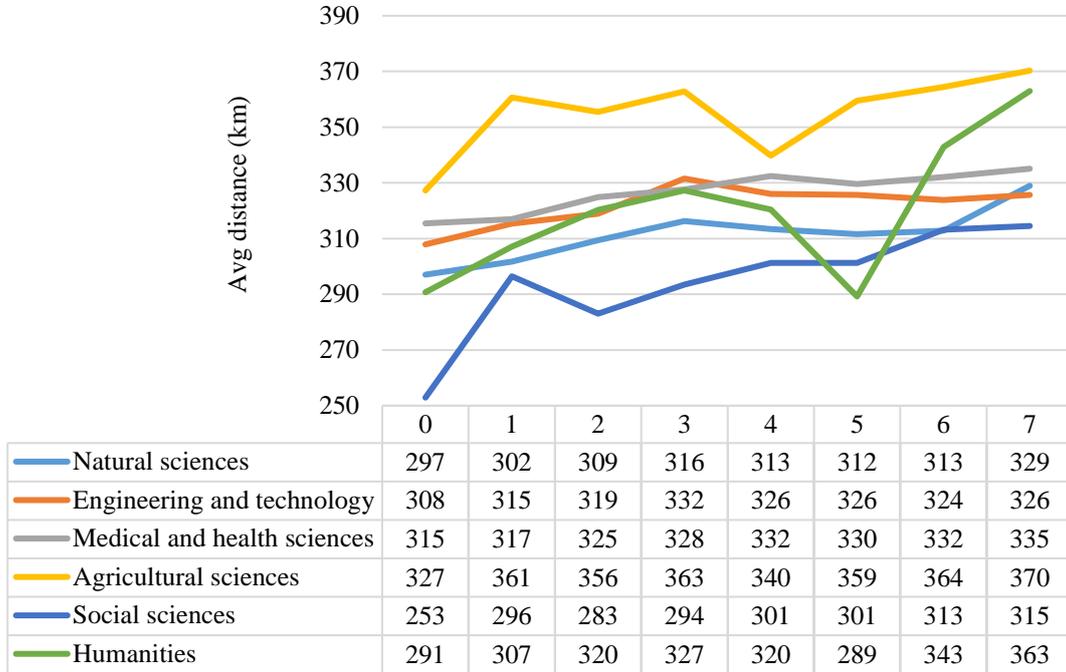

Citation delay: time elapsed between citing and cited publications dates (years)

*Figure 9: Gravity model applied to national flows: coefficient of distance with variation of time (citation window): dataset without self-citations*

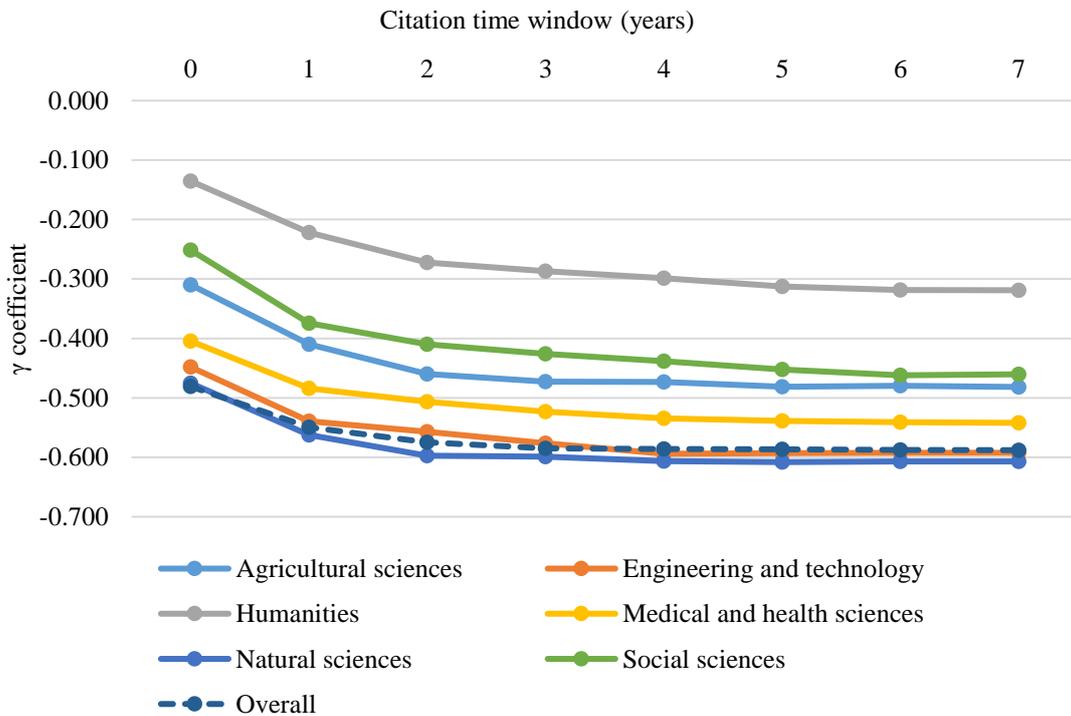

## 4. Conclusions

In Italy, as shown in other contexts, the factor of geographic proximity favors knowledge flows (Abramo, D'Angelo, & Di Costa, 2020a), including when accounted for the cognitive proximity of research in the territories concerned (Abramo, D'Angelo, & Di Costa, 2020b). The phenomenon is shown to be substantial in the national context, and to a lesser extent for Europe, but is scarcely observed in the transcontinental flows of knowledge. In this work we have investigated the sensitivity of the geographic proximity effect to the time passed since publication of the cited article, with the analyses both including and excluding self-citations.

The analysis on national flows shows that there is an increase over time in the average distance between the knowledge producers and users. The same analysis conducted with the exclusion of self-citations shows like trends, for all DAs, but with very significant increases in distances. This phenomenon is especially marked in the first years after publication, due to the habit of researchers to self-cite their recent works (Aksnes, 2003). In the comparison of the results "with" vs "without self-citations", the largest percentage changes in the average distances of proximity effect are in Engineering and technology, in all six years beginning from publication, and in Natural sciences in the seventh year.

The same type of analysis conducted on a European scale shows similar trends after publication, however for all DAs and at any time, there is significantly less difference in average distance between the with and without self-citations datasets. Instead, for transcontinental knowledge flows, the time patterns of average distance between producer and user seem more unstable, with trends even towards decreasing average distance beginning 3-4 years from publication. As well, similarly to the EUR case, for all DAs and time intervals, there is little effect from self-citations.

Even after accounting for the research sizes and relevant cognitive proximity of the territories, national flows are still affected by geographic proximity in all DAs, with impact decaying over time. As in the national case, although with more erratic trends, the same analysis of distance bias at the continental level shows that it again decreases over time.

This work also evidences that geographic bias tends to be differentiated across DAs, by virtue of their intrinsic characteristics. Humanities and Social sciences have a smaller area of influence, a smaller average range of citation flows; at the same time, the decay of citation flows with geographic distance is lower than in other DAs, particularly compared to the Sciences. This could be due to the peculiarity of the research topics addressed, more country specific for Humanities and Social sciences, and therefore with more localized spillovers, but also with lower citability of the works, as well as lower incidence of self-citations for these DAs compared to those of Sciences.

When we also exclude self-citations, the effect of geographic proximity is generally attenuated, and especially in DAs with a high citation intensity, i.e. probably with substantial shares of self-citations. The attenuation (with/without self-citations) is very strong near the date of publication, and then stabilizes, explained by the tendency of researchers to self-cite more recent works. Indeed all citing publications will tend to cite recent research, but this is even more the case when self-citing. Self-citation can be a tactical tool aimed to fostering visibility and asserting scientific authority, and can promote external citations (Van Raan, 2008). The more one cites oneself the more one is cited by other scholars. Controlling for numerous sources of variation in cumulative citations from others, Aksnes (2003) suggests that "each additional self-citation increases the number of citations from others by about one after one year, and by about three after five years".

All typical limits of bibliometric analyses apply to this work, in particular: i) publications are not representative of all knowledge produced; ii) bibliographic repertories do not index all journals and so not all publications; iii) not all citations are positive or indicate real use; and iv) citations are not representative of all uses.

Caution is also due in generalizing the results of any study based on country-specific data: in the matters at hand, it would be useful to develop comparisons with countries in various other global positions.